\documentclass[article,11pt,onecolumn,superscriptaddress]{revtex4-2}
\usepackage{graphicx}
\usepackage{amsmath,amsfonts}
\usepackage{color}
\usepackage{xcolor}
\usepackage{soul}
\usepackage[normalem]{ulem}
\usepackage{braket}
\newcommand{\upsub}[1]{\sb{\mathrm{#1}}}
\newcommand{\upsup}[1]{\sp{\mathrm{#1}}}

\usepackage{xr}
\makeatletter
\newcommand*{\addFileDependency}[1]{% argument=file name and extension
  \typeout{(#1)}
  \@addtofilelist{#1}
  \IfFileExists{#1}{}{\typeout{No file #1.}}
}
\makeatother

\newcommand*{\myexternaldocument}[1]{%
    \externaldocument{#1}%
    \addFileDependency{#1.tex}%
    \addFileDependency{#1.aux}%
}

\myexternaldocument{3R_SI}

\begingroup\lccode`~=`_\lowercase{\endgroup\let~\upsub}
\begingroup\lccode`~=`^\lowercase{\endgroup\let~\upsup}

\AtBeginDocument{%
	\catcode`_=12 \catcode`^=12
	\mathcode`_="8000
	\mathcode`^="8000
}

\begin{document}
\title{Ultrafast all-optical switching in nonlinear 3R-MoS$_2$ van der Waals metasurfaces}
	
\author{Levin Seidt}
\affiliation{Chair in Hybrid Nanosystems, Nanoinstitute Munich, Faculty of Physics, Ludwig-Maximilians-Universit{\"a}t M{\"u}nchen, 80539 Munich, Germany}	

\author{Thomas Weber}
\affiliation{Chair in Hybrid Nanosystems, Nanoinstitute Munich, Faculty of Physics, Ludwig-Maximilians-Universit{\"a}t M{\"u}nchen, 80539 Munich, Germany}	

\author{Albert A. Seredin}
\affiliation{School of Physics and Engineering, ITMO University, St. Petersburg 197101, Russia}

\author{Thomas Possmayer}
\affiliation{Chair in Hybrid Nanosystems, Nanoinstitute Munich, Faculty of Physics, Ludwig-Maximilians-Universit{\"a}t M{\"u}nchen, 80539 Munich, Germany}	

\author{Roman Savelev}
\affiliation{School of Physics and Engineering, ITMO University, St. Petersburg 197101, Russia}	

\author{Mihail A. Petrov}
\affiliation{School of Physics and Engineering, ITMO University, St. Petersburg 197101, Russia}

\author{Stefan A. Maier}
	\affiliation{School of Physics and Astronomy, Monash University, Clayton, Victoria 3800, Australia}
	\affiliation{The Blackett Laboratory, Department of Physics, Imperial College London, London, SW7 2BW, United Kingdom}

\author{Andreas Tittl}
\affiliation{Chair in Hybrid Nanosystems, Nanoinstitute Munich, Faculty of Physics, Ludwig-Maximilians-Universit{\"a}t M{\"u}nchen, 80539 Munich, Germany}

\author{Leonardo de S. Menezes}
\email{l.menezes@physik.uni-muenchen.de}
\affiliation{Chair in Hybrid Nanosystems, Nanoinstitute Munich, Faculty of Physics, Ludwig-Maximilians-Universit{\"a}t M{\"u}nchen, 80539 Munich, Germany}
\affiliation{Departamento de F\'{i}sica, Universidade Federal de Pernambuco, 50670-901 Recife-PE, Brazil}

\author{Luca Sortino}
\email{luca.sortino@physik.uni-muenchen.de}
\affiliation{Chair in Hybrid Nanosystems, Nanoinstitute Munich, Faculty of Physics, Ludwig-Maximilians-Universit{\"a}t M{\"u}nchen, 80539 Munich, Germany}
\date{\today}
\maketitle

\noindent
\textbf{Second-order nonlinear optical processes are fundamental to photonics, spectroscopy, and information technologies, with material platforms playing a pivotal role in advancing these applications. Here, we demonstrate the exceptional nonlinear optical properties of the van der Waals crystal 3R-MoS$_2$, a rhombohedral polymorph exhibiting high second-order optical susceptibility ($\chi^{(2)}$) and remarkable second-harmonic generation (SHG) capabilities. By designing high quality factor resonances in 3R-MoS$_2$ metasurfaces supporting quasi-bound states in the continuum (qBIC), we first demonstrate SHG efficiency enhancement exceeding 10$^2$. Additionally, by using degenerate pump-probe spectroscopy, we harness the \textit{C}$_{3v}$ system's symmetry to realize ultrafast SHG polarization switching with near-unity modulation depth. The operation speeds are limited only by the excitation pulse duration, allowing its characterization via the nonlinear autocorrelation function. These findings establish 3R-MoS$_2$ as a transformative platform for nanoscale nonlinear optics, offering large conversion efficiencies and ultrafast response times for advanced pulse measurement devices, integrated photonics, and quantum technologies.}

\newpage
\pagebreak

\noindent
\textbf{Introduction}

\noindent
The class of van der Waals (vdW) materials, particularly in the form of atomically thin two-dimensional (2D) layers, has emerged as a versatile platform for nanoscale optical devices \cite{li2024optical}. Their exceptional linear and nonlinear optical properties, combined with substrate affinity and engineering of layers stacking order, make them highly attractive for applications in integrated photonics \cite{meng2023photonic}.
Among nonlinear processes, second-order effects such as second harmonic generation (SHG) have been extensively used in 2D materials for structural and optical characterization \cite{Mennel2018,lafeta2024probing}, frequency conversion \cite{autere2018nonlinear}, and all-optical modulation \cite{klimmer2021}. 
More recently, the optical properties of multilayered films of transition metal dichalcogenides (TMDCs) have sparked interest in the nanophotonics community owing to their high refractive indexes \cite{Munkhbat2022}, large optical anisotropy \cite{Ermolaev2020} and deterministic manipulation of material layers \cite{voronin2024chiral}. These properties opened extensive studies on the use of TMDCs as building blocks of optically resonant nanostructures \cite{Lin2022,Munkhbat2022b,Zotev2022b}. Several structures,  such as waveguides \cite{ling2023}, optical resonators \cite{Verre2018}, and metasurfaces \cite{nauman2021tunable,weber2023,sortino2024van} have been realized directly from exfoliated TMDCs multilayers. 
Moreover, the realization of lasing regime \cite{sung2022room}, strong light-matter coupling \cite{weber2023,sortino2024van}, and spontaneous parametric down-conversion (SPDC) \cite{trovatello2025} in photonically engineered multilayered vdW materials hold promise for transforming the design principles of nanophotonic architectures.

However, the common 2H phase of multilayer TMDCs is inversion symmetric \cite{wagoner1998}, canceling second order nonlinear responses and limiting opportunities for more practical photonic devices based on vdW materials. At the same time, the overall SHG efficiencies in monolayer TMDCs remain low, due to their atomic thickness and the limited interaction time with the excitation field \cite{trovatello2021,khurgin2023}, posing a limit to practical applications.
Early studies on SHG in TMDC polymorphs revealed the rhombohedral 3R phase as a promising platform for nonlinear optics \cite{strachan2021}, due to its exceptionally large second-order nonlinear susceptibility ($\chi^{(2)} > 400$ pm/V), significantly exceeding that of the 2H phase \cite{wagoner1998,zhao2016}. This layer-independent nonlinearity has positioned 3R-TMDCs at a focal point for nonlinear optics \cite{shi2017} and rising candidates for nanophotonic applications based on vdW materials \cite{trovatello2024tunable}.
Mechanically exfoliated 3R-TMDCs exhibit exceptionally high SHG efficiencies, up to $0.03\%$ via periodic poling \cite{trovatello2025}, rivaling or surpassing established nonlinear materials \cite{xu2022}. Additionally, the process of SPDC has been demonstrated, enabling entangled photon pair generation for quantum optics applications \cite{weissflog2024,feng2024polarization,trovatello2025}. Furthermore, the high refractive index allows for efficient second harmonic (SH) waveguiding on lower-index substrates \cite{xu2025spatiotemporal}, while advancements in layer engineering have enabled novel approaches to twist-engineer thephase matching, leveraging the unique properties of vdW materials \cite{tang2024quasi,trovatello2025}. These properties, togheter with emerging capabilities for large-area growth \cite{qin2024}, place 3R-TMDCs as a scalable and versatile platform for nonlinear photonics.

A key strategy to further enhance the nonlinear optical response of dielectric materials, beyond relying solely on their intrinsic properties, is the engineering of multipolar optical resonances through nanostructuring into single nanoparticles or metasurfaces \cite{cortes2022optical,Lin2022}. This approach increases the local electromagnetic field amplitude and interaction time, leading to significant enhancements in both linear and nonlinear optical properties \cite{kuznetsov2016optically}. In particular, nanophotonic engineering has been successfully applied to exfoliated 3R-TMDCs, resulting in SHG enhancements in individual nanodisks \cite{zograf2024}, periodic arrays \cite{ling2024nmeta}, and guided modes in two-dimensional metasurfaces \cite{zograf2024meta}. 
While providing ways to increase material nonlinearities, these nanophotonic approaches lacks tunability as they are restricted to geometrical resonances introduced in the material layer. A novel approach leverages symmetry-broken metasurfaces based on quasi-bound states in the continuum (qBICs) \cite{koshelev2018asymmetric} which provide a widely tunable and highly efficient framework for enhancing nonlinear optical interactions. By supporting high quality ($Q$) factor resonances with wide spectral tunability, qBIC metasurfaces enable strong field confinement and selective SHG enhancement at resonance, making them an ideal platform for nonlinear photonics applications \cite{liu2019high,kang2023applications}.
Furthermore, the symmetry properties of 3R-TMDCs further shape their nonlinear response. While the in-plane crystal symmetry of 3R-MoS$_2$ ($C_{3v}$) is analogous to that of monolayers ($D_{3h}$), the absence of a horizontal mirror plane ($\sigma_h$) allows for an out-of-plane SHG component. Ultrafast SHG control has been previously demonstrated in $D_{3h}$ monolayers \cite{klimmer2021}, yet all-optical modulation in $C_{3v}$ systems remains unexplored.  Despite significant advancements in TMDC-based nanophotonics for light-matter interaction applications, the role of qBICs in enabling ultrafast optical control of nonlinear responses in 3R-TMDC metasurfaces remains largely unaddressed.

In this work, we combine the intrinsically large second-order response of 3R-MoS$_2$ with the rational design of qBIC dielectric metasurfaces with high-$Q$ factors ($Q>100$), substantially boosting the nonlinear optical conversion efficiency.  
First-principles modeling of the metasurface nonlinear properties closely aligns with experimental results, confirming enhancement factors exceeding $10^3$ and that the qBIC-driven SHG enhancement is largely independent of the crystal orientation. In addition, by leveraging the $C_{3v}$ symmetry of 3R-MoS$_2$ we characterize ultrafast laser pulses, as a function of time and frequency. Employing degenerate pump-probe spectroscopy, we demonstrate ultrafast all optical modulation and polarization switching of the SH output and perform nonlinear autocorrelation measurements of pulse durations.
Together, these findings bridge the gap between high-speed modulation and high nonlinear conversion efficiency, advancing practical implementations of integrated vdW photonics, including ultrafast optical switches, compact frequency converters, and advanced pulse measurement technologies.

\begin{figure}[b]
    \centering
    \includegraphics[width=1\linewidth]{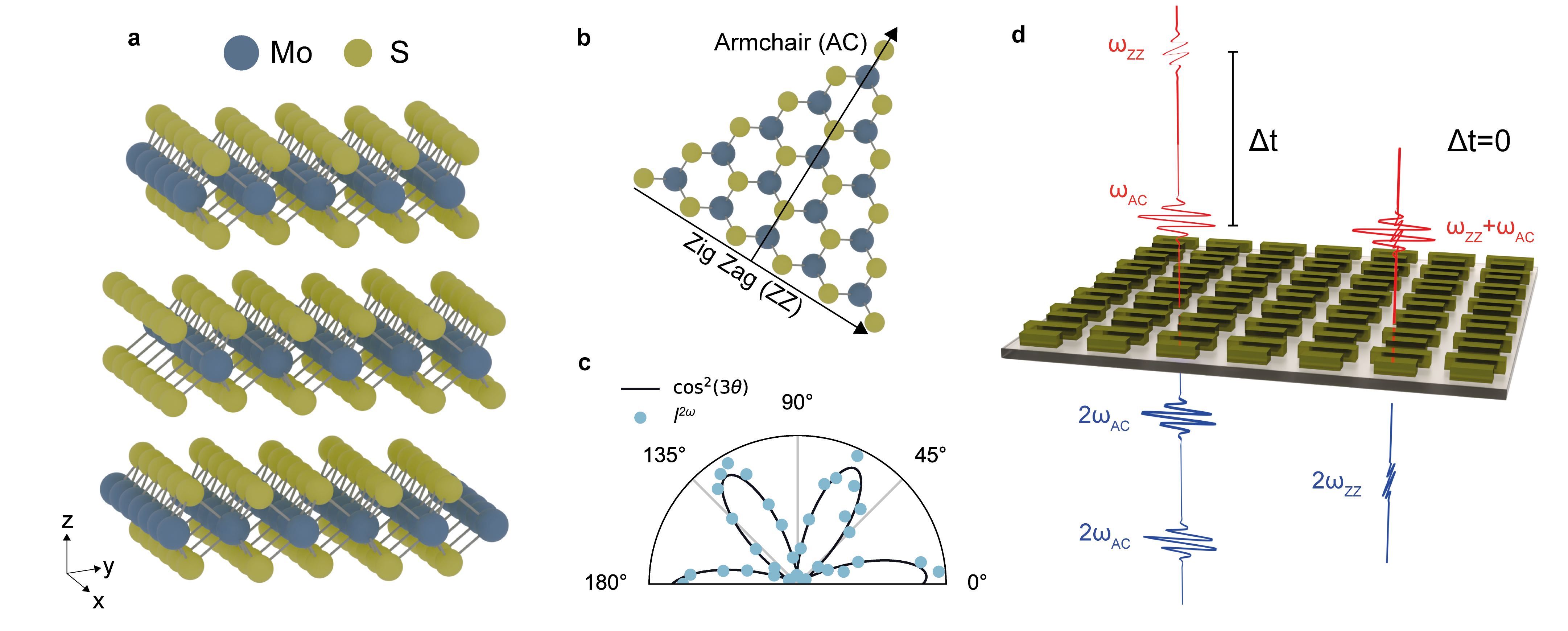}
    \caption{\textbf{Crystalline symmetry and SHG optical selection rules in 3R-MoS$_2$ metasurfaces.} 
    (\textbf{a}) Crystalline structure of 3R-MoS$_2$, showing the vertical stacking of three MoS$_2$ monolayers in an ABC rhombohedral arrangement. Molybdenum (Mo) atoms are depicted in blue, sulfur (S) atoms in yellow. 
    (\textbf{b}) In-plane $D_{3h}$ crystal symmetry of a single MoS$_2$ layer, with the zigzag (ZZ) and armchair (AC) crystallographic directions indicated by black arrows. 
    (\textbf{c}) Polar plot of the SHG intensity, $I^{2\omega}$, for a bulk 3R-MoS$_2$ crystal, measured by rotating both the excitation and collection linear polarization angle. The solid line represents a fit to a $\approx\cos^2(3\theta)$ function, confirming the six-fold symmetry of the crystal and the preferential nonlinear emission along the AC axis, oriented approximately at 0 degrees. 
    (\textbf{d}) Schematic illustration of a symmetry-broken 3R-MoS$_2$ metasurface supporting quasi-bound states in the continuum for resonant SHG enhancement, along with the SHG optical selection rules for a degenerate pump-probe excitation scheme. To achieve SHG optical switching, two fundamental pulses, orthogonally polarized along the AC ($\omega_{AC}$) and ZZ ($\omega_{ZZ}$) axes, excite the sample at normal incidence. When the time delay ($\Delta t$) between the pulses exceeds their duration, SH emission is polarized along the AC axis ($2\omega_{AC}$). When both pulses overlap in time, the SH signal is instead polarized along the ZZ axis ($2\omega_{ZZ}$). For clarity, we have neglected the additional components of the SHG outputs to better illustrate the switching mechanism. }
    \label{fig1}
\end{figure}

\bigskip
\noindent
\textbf{Results}

\noindent
\textbf{SHG optical selection rules in 3R-TMDCs} 
The rhombohedral 3R and hexagonal 2H phases of TMDCs differ significantly in their stacking arrangements, which profoundly impact their symmetry and nonlinear optical properties. 
In the more common 2H phase, adjacent atomic layers are rotated by 180 degrees relative to each other, resulting in an inversion-symmetric structure that cancels second-order nonlinear optical responses \cite{wagoner1998}. 
In contrast, the 3R phase features an ABC-stacking arrangement, where all atomic layers are oriented in the same direction, breaking inversion symmetry even in multilayered samples (Figure \ref{fig1}a).
The SHG in TMDCs monolayers is determined by the $D_{3h}$ point group which allows the non vanishing components of the $\chi^{(2)}$ nonlinear tensor \cite{malard2013}: $\chi^{(2)}_{yyy} = -\chi^{(2)}_{yxx} = -\chi^{(2)}_{xxy} = -\chi^{(2)}_{xyx}$, where $y$ corresponds to the armchair (AC) direction and $x$ to the zig-zag (ZZ) direction (Figure \ref{fig1}b). 
Consequently, the out-of-plane nonlinear polarization at the SH frequency vanishes, $P_z(2\omega)=0$. 
In multilayered 3R-TMDCs, however, the rhombohedral stacking of the bulk 3R-phase removes the horizontal mirror plane symmetry ($\sigma_h$) and reduces the symmetry from $D_{3h}$ to the $C_{3v}$ point group, similar to GaAs or LiNbO. 
This leads to additional independent nonzero elements of the $\chi^{(2)}$ tensor: $\chi^{(2)}_{zzz}, \quad \chi^{(2)}_{xzx} = \chi^{(2)}_{yzy}, \quad \chi^{(2)}_{xxz} = \chi^{(2)}_{yyz}, \quad \chi^{(2)}_{zxx} = \chi^{(2)}_{zyy}$, introducing a nonzero contribution from $P_z(2\omega)$.

Under normal incidence excitation normal component of the field vanishes, $E_{z} = 0$, and the SHG selection rules of the unstructured layer of 3R-TMDCs reduce to the monolayer case~\cite{klimmer2021}, with the SH signal exhibiting characteristic six-fold rotational symmetry (Figure \ref{fig1}c). 
In the case of metasurfaces considered in this work all components of the total electric field induced inside the nanoresonators are generally non-zero. However, further symmetry analysis and numerical simulations (see Methods and SI) reveal that nonlinear tensor corresponding to the $D_{3h}$ point group is sufficient to describe SHG process in metasurfaces under excitation by a normally incident plane wave. This is mostly due to much higher nonlinear susceptibility of the $D_{3h}$ group compared to other elements of the $C_{3v}$ nonlinear tensor in 3R-TMDCs~\cite{zograf2024}, which provide only negligible contribution to the overall SHG signal. The nonlinear polarization at the second harmonic frequency then can be simplified as follows:
\begin{eqnarray}
P_{ZZ}^{2\omega} &\propto& 2\,E_{AC}\,E_{ZZ}, \\
P_{AC}^{2\omega} &\propto& E_{AC}^2 - E_{ZZ}^2.
\end{eqnarray}
where $E_{AC}$ and $E_{ZZ}$ represent the excitation electric field amplitudes along the AC and ZZ direction, respectively (see also Methods). 
Note that when we excite the metasurface with a plane wave polarized either along the AC or ZZ direction, both $P_x^{2\omega}$ and $P_y^{2\omega}$ components are nonzero. However, they both emit SH light polarized along the AC direction, resulting in $I_{AC}^{2\omega} \ne 0$ and $I_{ZZ}^{2\omega} = 0$. To generate SH emission polarized along the ZZ axis ($I_{ZZ}^{2\omega} \ne 0$), both fundamental field components ($E_{AC} \ne 0, E_{ZZ} \ne 0$) must be present. 
Such all-optical modulation of the 3R-MoS$_2$ SHG signal polarization can be realized by combining two orthogonally polarized laser pulses at normal incidence~\cite{klimmer2021}. This principle is schematically illustrated in Figure \ref{fig1}d, where two collinear pulses at the fundamental frequency ($\omega$) interact within a nanostructured 3R-MoS$_2$ qBIC metasurface at normal incidence. Initially, the resulting SH signal ($2\omega$) is polarized along the AC direction. However, at zero time delay, when the two fields overlap, the SH polarization acquires a nonzero ZZ-polarized component. This all-optical SHG modulation mechanism, governed by the interplay between fundamental field symmetry and nonlinear polarization, enables ultrafast polarization switching of the SH signal in 3R-MoS$_2$ metasurfaces on a timescale limited only by the fundamental pulse temporal width. 

\begin{figure}
\centering
\includegraphics[width=0.5\linewidth]{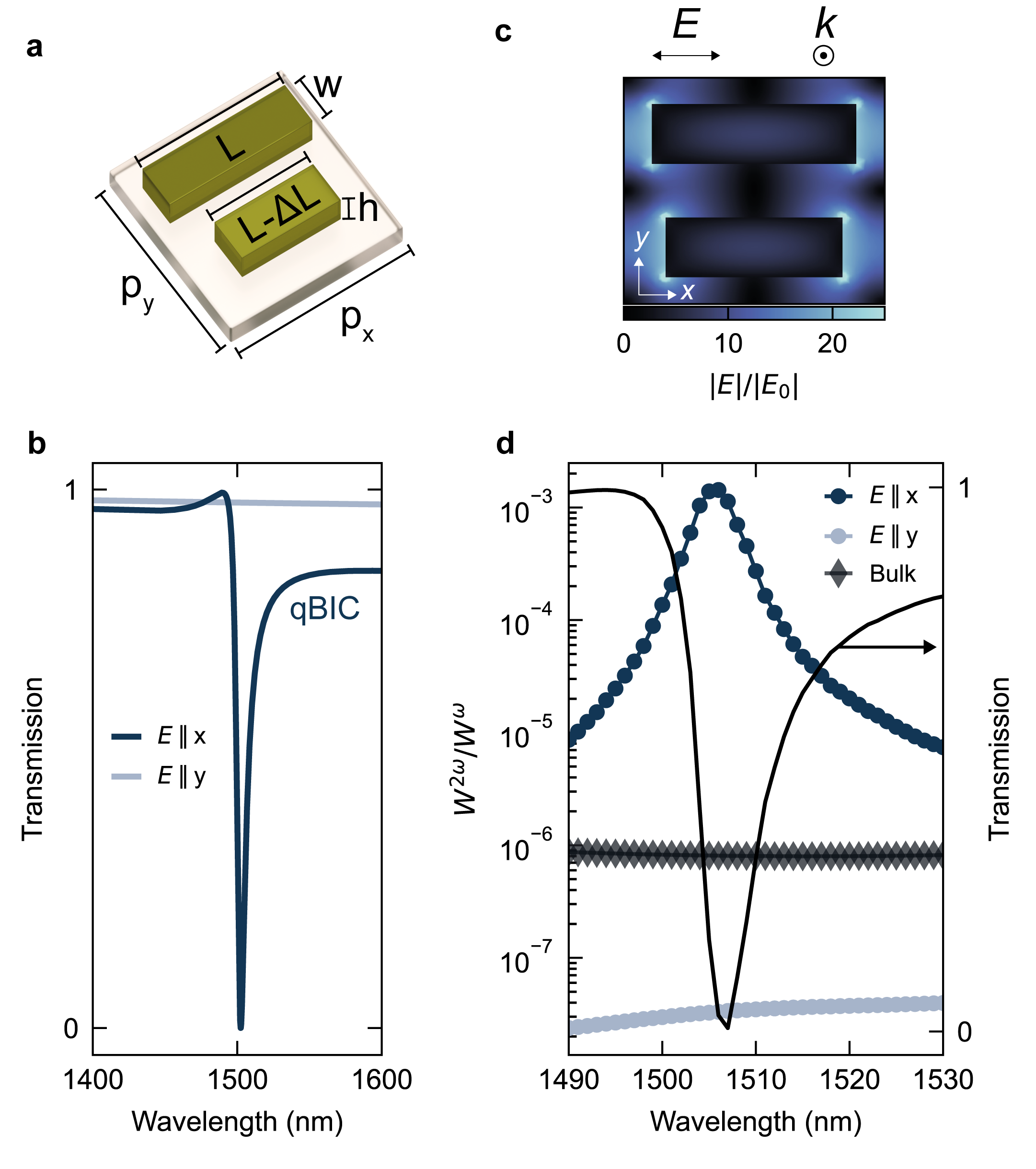}
\caption{ 
\textbf{3R-MoS$_2$ metasurface design for qBIC-driven second-order nonlinear enhancement} 
(\textbf{a}) Schematic illustration of the unit cell of the symmetry-broken qBIC metasurface, independent on the crystal axis orientation. The symmetry-breaking condition, introduced by a length difference ($\Delta L$), gives rise to the radiative qBIC state. 
(\textbf{b}) Simulated optical transmission spectra of a 3R-MoS$_2$ qBIC metasurface for linearly polarized excitation parallel ($E \|x$) or perpendicular ($E\|y$) to the long axis of the nanorords.
(\textbf{c}) Spatial distribution of the simulated electric field ($E$) normalized to the incident field ($E_0$), for excitation polarized parallel to the nanorods, along the x-axis, and calculated over a plane at half-height of the resonators. 
(\textbf{d}) SHG power $W^{2\omega}$, normalized to the fundamental excitation power $W^{\omega}$, for a bulk crystal (diamond markers) and a qBIC metasurface (circular markers), excited either parallel or perpendicular to the nanorods, both with a thickness of 218 nm. The qBIC metasurface is tuned for a wavelength around 1500 nm: the black solid line depicts its optical transmission and is spectrally correlated with the SHG enhancement.
}
\label{fig2}
\end{figure}

\smallskip
\noindent
\textbf{Design principles for SHG enhancement in qBIC metasurfaces} 
By leveraging the strong electric fields and precise resonance control provided by qBIC states in optical metasurfaces, enhancement of nonlinear conversion efficiencies can be achieved \cite{anthur2020continuous}.  The symmetry-broken qBIC metasurface unit cell design used in our work is illustrated in Figure \ref{fig2}a. 
Each unit cell consists of two nanorods, where the length difference ($\Delta L$) introduces a symmetry-breaking condition that facilitates the generation of a bright qBIC resonance \cite{koshelev2018asymmetric}. 
%The rod design is chosen as it provides a single resonance within a broad spectral range, highly dependent on the excitation of the nanostructures. 
The symmetry-breaking mechanism allows for the realization of qBIC resonances with high $Q$ factors by simply tuning the $\Delta L$ value \cite{liu2019high}, limited solely by fabrication and material quality. In addition, the qBIC wavelength can be precisely controlled by rescaling all unit cell parameters by a scaling factor ($S$) \cite{kuhner2023high}, enabling an additional parameter to fine-tune the resonance spectral position for tailored nonlinear and quantum optical applications. 

We engineer the qBIC resonance in a fabricated 3R-MoS$_2$ metasurface to be centered at 1500 nm, ensuring that the generated SH signal wavelength remains above the optical absorption region of intrinsic exciton resonances, below 700 nm \cite{Zotev2022b}. The unit cell is tuned to an arbitrary value of $S=1.10$ with the corresponding geometric parameters: length $L$ = 597 nm, asymmetry $\Delta L$ = 79 nm, periodicity $p_y$ = $p_x$ = 763 nm, width $w$ = 202 nm, and height $h$ = 218 nm. Figure \ref{fig2}b shows the numerical simulations for the optical transmission of a qBIC metasurface in a 218 nm-thick 3R-MoS$_2$ film, where the excitation light is linearly polarized either parallel or perpendicular to the long axis of the nanorods, corresponding to the \textit{x}- and \textit{y}-axis, respectively. When the excitation is polarized parallel to the nanorods, the genuine qBIC mode at 1500 nm is excited (Figure \ref{fig2}c), while for excitation perpendicular to the nanorods the resonance vanishes. This effect is related to the large anisotropy of the unit cell's symmetry, which plays a predominant role in maximizing the nonlinear optical response of designed 3R-MoS$_2$ metasurfaces.

Finally, we model the nonlinear response with numerical simulations (see Methods for more details), computing the SHG emission for both a qBIC metasurface and a reference 3R-MoS$_2$ film of identical thickness. The solid black line in Figure $\ref{fig2}$d represents the numerically computed optical transmission spectrum of the metasurface supporting a qBIC mode at 1500 nm, depicted for a linearly polarized excitation at the fundamental frequency, with the electric field aligned along the nanorods. Under this type of excitation, a SH signal is generated (indicated by blue circular markers in Figure \ref{fig2}d). Here, the computed SH power, $W^{2\omega}$, is normalized to the power of the excitation at the fundamental frequency, $W^\omega$. For comparison, the graph also depicts SHG spectra from the qBIC metasurface under non-resonant excitation (gray circular markers) as well as from a bulk 3R-MoS$_2$ film of identical thickness (diamond markers). At the qBIC resonance, the metasurface achieves an SHG efficiency enhancement of approximately $10^3$ compared to the bulk film. This enhancement arises from the strong local field confinement associated with the qBIC mode. In contrast, under non-resonant (orthogonal) excitation, the SHG efficiency falls below that of the bulk film. This decrease is attributed to interference effects between the resonant fields within the nanostructured sample, which effectively suppress the SHG output.

\begin{figure}[b]
\centering
\includegraphics[width=1\linewidth]{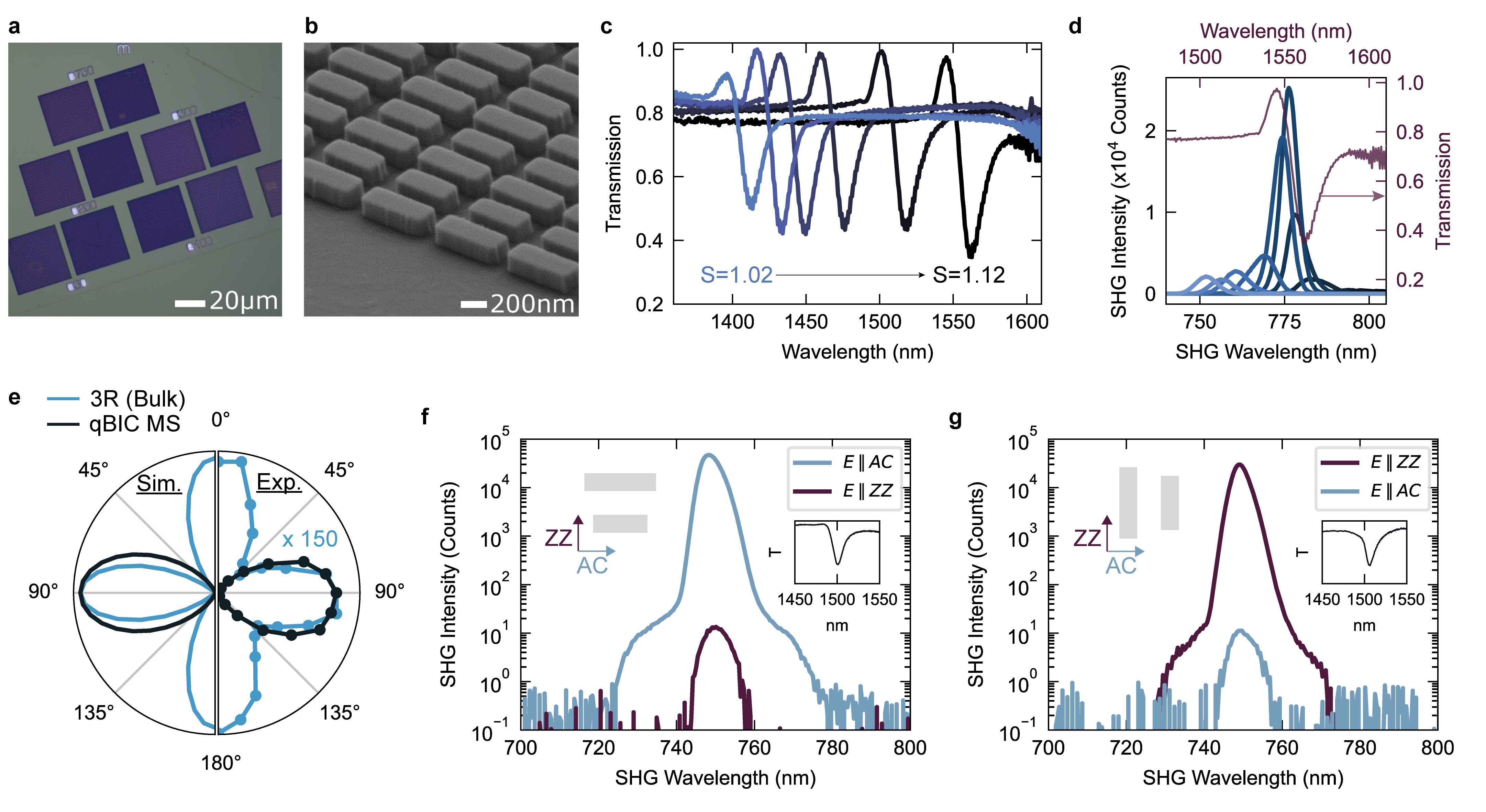}
\caption[]{ \textbf{Second-order nonlinear optical response of 3R-MoS$_2$ metasurfaces.} 
(\textbf{a}) Bright-field optical image of a fabricated qBIC metasurface from an exfoliated 3R-MoS$_2$ crystal. 
(\textbf{b}) Tilted scanning electron microscope (SEM) image of a 3R-MoS$_2$ qBIC metasurface on a glass substrate.  
(\textbf{c}) Normalized white-light transmission spectra of 3R-MoS$_2$ qBIC metasurfaces (thickness 218 nm) for varying scaling factors ($S$) from 1.02 to 1.12, showing a redshift of the qBIC resonance with increasing unit cell size. 
(\textbf{d}) SHG intensity as a function of the fundamental excitation wavelength, scanned around the qBIC resonance for a metasurface ($S=1.12$, in purple the metasurface optical transmission spectrum).
(\textbf{e}) Polar plot of the SHG intensity as a function of the excitation beam polarization angle, shown for a 3R-MoS$_2$ qBIC metasurface (MS, $S=1.10$ and 218 nm thickness) compared to a bulk reference sample with the same height. Left: normalized SHG signal obtained from the the nonlinear tensor model (see Methods and Supplementary Note II). Right: experimental SHG signal from a qBIC metasurface (as in Figure \ref{fig3}f) and a reference bulk 3R-MoS$_2$ multiplied 150 times. The signal is collected by rotating the excitation laser beam, with a fixed linear polarizer in the collection path along AC.
(\textbf{f,g}) SHG spectra for excitation polarized parallel or perpendicular to the unit cell nanorods, for metasurfaces aligned along the AC (f) and ZZ (g) crystal directions, both showing strong SHG enhancement at the qBIC resonance centered at 1500 nm. Inset: optical transmission of each relative metasurface, showing the well-resolved qBIC resonance at approximately 1500 nm.
}
\label{fig3}
\end{figure}

\smallskip
\noindent
\textbf{SHG enhancement in 3R-MoS$_2$ qBIC metasurfaces}
The designed qBIC metasurfaces are fabricated from exfoliated 3R-MoS$_2$ layers with 218 nm thickness following a standard top-down nanofabrication process \cite{weber2023}. An optical micrograph of a fabricated metasurfaces array (Figure \ref{fig3}a) shows well-defined nanostructured regions, where the excess material is removed via chemical etching. Scanning electron microscopy confirms the high uniformity and quality of the final metasurfaces (Figure \ref{fig3}b). To tune the qBIC mode around 1500 nm, we fabricated metasurfaces with different scaling factors, $S$, from the same exfoliated sample. The white light optical transmission of the fabricated array reveals the systematic shift of the qBIC resonance wavelength as a function of the $S$ values (Figure \ref{fig3}c), while retaining a high Q factor of over 120 (Supplementary Figure S\ref{fig:SI-tcmt-fit}). The impact of the qBIC resonance on nonlinear optical conversion is demonstrated in wavelength-dependent SHG measurements from a metasurface sample ($S=1.12$), where the SHG intensity exhibits a strong enhancement around 775 nm, aligning with the fundamental qBIC mode at 1550 nm (Figure \ref{fig3}d). 

The polarization dependence of the SHG response provides further insights into the role of the system's symmetry in the nonlinear process. Figure \ref{fig3}e shows a comparison of the SHG power obtained from the numerical model (left) and the collected SHG intensity in experiments (right), shown for both bulk 3R-MoS$_2$ and a qBIC metasurface ($S=1.10$) of the same thickness, and with nanorods aligned along the AC axis. The SHG signal is analyzed as a function of the polarization angle of the incident excitation pulse at 1500 nm, with a fixed linear polarizer in the collection path aligned along AC. The bulk response reveals the expected maximal emission from the AC crystallographic axes with a 4-fold symmetry \cite{psilodimitrakopoulos2018ultrahigh}. On the other hand, the modified SHG emission pattern due to symmetry-breaking in qBIC metasurfaces shows a predominant emission along the long axis of the nanorods, where the qBIC field is maximized. The observed SHG polar pattern is in excellent agreement with the nonlinear model, as shown in Supplementary Note II.

To further investigate the influence of the crystal orientation on the SHG enhancement, we compare two fabricated metasurfaces where the nanorods are aligned either along the AC or ZZ directions, with resonances centered around 1500 nm (Figures \ref{fig3}f,g). The insets in Figures \ref{fig3}f,g confirm the presence of the qBIC mode through the corresponding transmission dip at 1500 nm. The SHG spectra were then acquired for fundamental excitation polarized along both crystallographic axes, with no polarization optics in the collection path. 
In the sample with nanorods oriented along the AC axis (Figure \ref{fig3}f) we observe a three-orders-of-magnitude SHG enhancement when switching the excitation field polarization from $E\|$ZZ to $E\|$AC. Similarly, for the sample with nanorods aligned along the ZZ axis (Figure \ref{fig3}g), a comparable SHG enhancement is observed for $E\|$ZZ, when the qBIC mode is efficiently excited. Despite the variation in the crystal alignment of the metasurface and polarization of the incident light, the emitted SHG signal remains directed along the AC axis, consistent with the selection rules previously described (Equations 1-2).
These observations demonstrate that the SHG enhancement above 10$^3$ is driven by the qBIC resonant fields, dictated by the orientation of the nanorods with respect to the polarization of the incident electric field, while remaining largely independent of the crystallographic orientation of the metasurface.

\begin{figure}[b]
\centering
\includegraphics[width=1\linewidth]{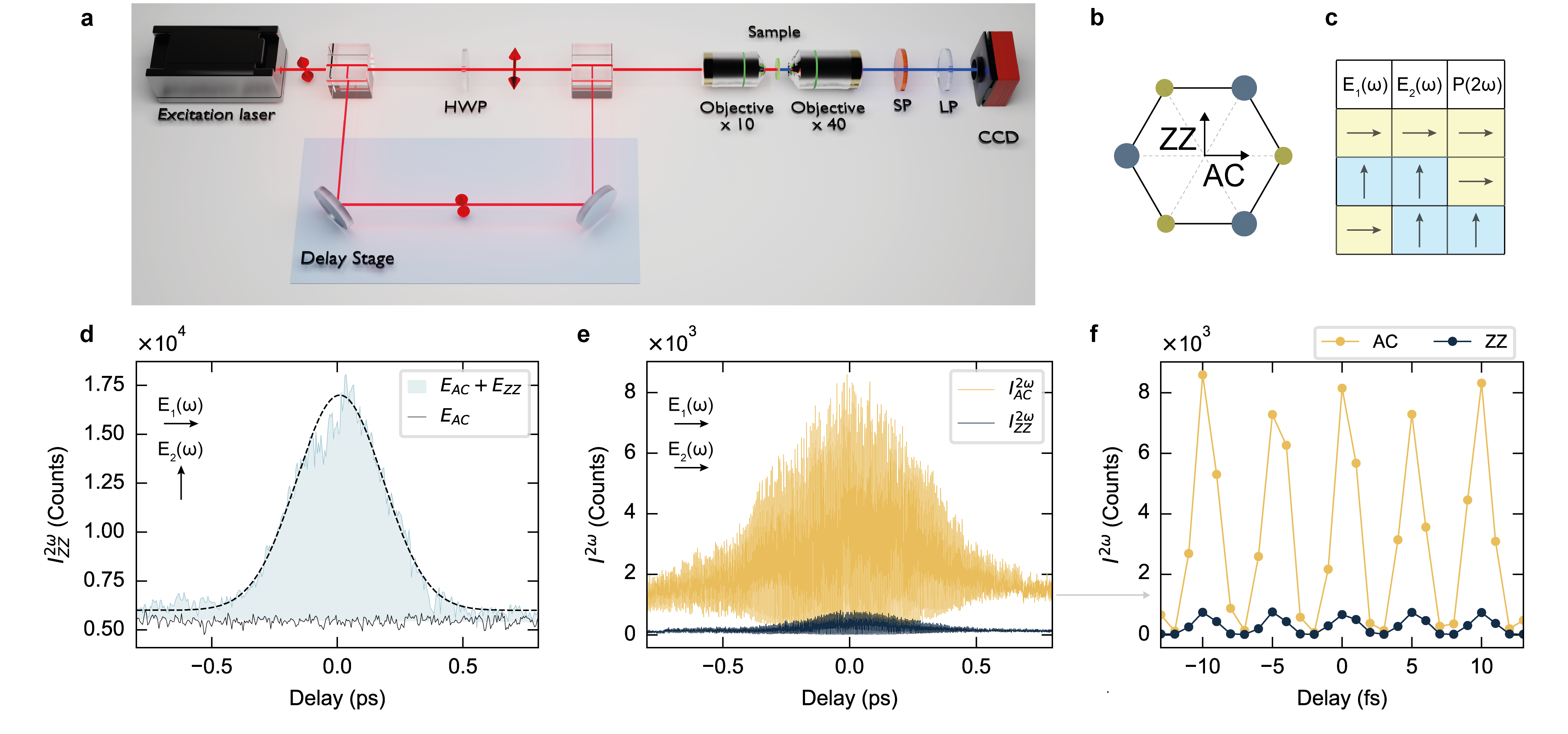}
\caption[]{\textbf{Ultrafast all-optical SHG switching and modulation in 3R-MoS$_2$ qBIC metasurfaces.} 
(\textbf{a}) Schematic of the ultrafast spectroscopy setup. The linearly polarized excitation laser is split into two arms, with one passing through a delay line and the other through a half-wave plate (HWP) to control the polarization of the second pulse. The pulses are then focused onto the sample in a confocal geometry, and the SHG signal is detected using a spectrometer (not shown) and a CCD camera after filtering (SP: shortpass filter,  LP: linear polarizer).
(\textbf{b}) In-plane structure of a 3R-MoS$_2$ monolayer crystal, indicating the armchair (AC) (horizontal arrows) or zigzag (ZZ) (vertical arrow) directions.
(\textbf{c}) Table of optical selection rules for excitation with two linearly polarized optical pulses, $E_1(\omega)$ and $E_2(\omega)$, and the resulting polarization direction if the SHG signal, $P(2\omega)$, at zero-time delay. The SHG is directed along the ZZ direction only in case of perpendicular polarization between the excitation pulses.
(\textbf{d}) SHG intensity from a qBIC metasurface ($S=1.10$) with linear polarization detection along the ZZ direction, $I_{ZZ}^{2\omega}$, as a function of the time delay between two excitation pulses, one polarized along AC and the other along ZZ (schematic inset). The data is compared with excitation with only one pulse along AC (black solid line). The dashed black line represents a Gaussian fit to the data.
(\textbf{e}) SHG intensity from a qBIC metasurface, aligned along the AC direction, as a function of the time delay between two excitation pulses polarized along AC (schematic inset). The SHG signal is recorded for linearly polarized detection along AC ($I_{AC}^{2\omega}$, in yellow) and along ZZ detection ($I_{ZZ}^{2\omega}$, in blue).
(\textbf{f}) SHG signal exhibiting a period of approximately 5 fs, originating from the interference of the two collinear beams along AC.}
\label{fig4}
\end{figure}

\smallskip
\noindent
\textbf{Ultrafast all-optical SHG polarization switching} 
We combine the SHG enhancement with the optical selection rules of 3R-MoS$_2$ to demonstrate qBIC metasurfaces for ultrafast control of the SH signal.
The degenerate pump-probe  setup used for the ultrafast spectroscopy experiments is depicted in Figure \ref{fig4}a. The excitation source generates linearly polarized pulses, which are split into two optical paths. One of these paths includes a half-wave plate to control the polarization angle of the second beam. The pulses are focused onto the sample using a low numerical aperture (NA) objective and collected with a high-NA objective in a confocal geometry to maximize signal collection efficiency. The generated SH signal is then analyzed using a CCD detector (see Methods for further details).

Figure \ref{fig4}b illustrates the optical selection rules governing 3R-MoS$_2$ SHG emission for different polarization configurations of the two pulses. 
When the pump and probe pulses are orthogonally polarized along the ZZ and AC directions (Figure \ref{fig4}c), the SHG emission undergoes polarization switching. 
Specifically, when the two pulses coincide temporally, the SHG signal shows a maximum along the ZZ polarization direction, effectively switching from the AC direction. 
This is shown in Figure \ref{fig4}d for a 3R-MoS$_2$ qBIC metasurface ($S=1.10$) oriented along the AC direction, and with qBIC resonance centered at 1500 nm. 
The maximized SHG signal detected along ZZ at zero-time delay confirms the effective ultrafast polarization switching mechanism. Furthermore, the time-dependent SHG peak can be used as an autocorrelation measurement between the two pulses, effectively giving the measurement of their duration. We fitted the data with a Gaussian function (dashed line in Figure \ref{fig4}d) and extracted a full-width at half-maximum (FWHM) of approximately 430 fs. 
This corresponds to an autocorrelation function of the optical parametric oscillator (OPO) pulses of $\approx280$ fs, in good agreement with the temporal dispersion induced by the optical components on the 200 fs pulses emitted at the OPO source output.

Next, we excited the metasurface with two parallel pulses along AC. The resulting SHG spectrum (Figure \ref{fig4}e and Supplementary Figure 3), measured as a function of the time delay between the two pulses, reveals the conventional interferometric autocorrelation. The interference between the two pulses strongly modulates the SHG signal, reaching near-unity modulation depth on a timescale of approximately 5 fs (Figure \ref{fig4}f), corresponding to the optical cycle at fundamental wavelength of 1500 nm.
We note the presence of a residual signal detected along the ZZ direction, $I_{ZZ}^{2\omega}\neq 0$, in contrast with the selection rules described in Equation 1-2, while in a reference unpatterned 3R-MoS$_2$ sample (Supplementary Figure 4) we detect negligible $I_{ZZ}^{2\omega}$ signal, as expected. The source of this polarization could be derived from the intrinsic modification of the nonlinear tensor and increased surface-to-volume ratio in the case of nanostructured metasurfaces, compared to planar thin films, or small misalignment in the angle between the crystal axis and the nanorod geometry.

\bigskip
\noindent
\textbf{Discussion}

\noindent
In conclusion, we have demonstrated large SHG efficiency enhancement in the vdW nonlinear material 3R-MoS$_2$ by rationally designing high-$Q$ factor resonances leveraging qBIC metasurfaces. In addition, the $C_{3v}$ symmetry of 3R-MoS$_2$ allows to achieve ultrafast all-optical switching and near unity modulation of the SHG signal, combining high efficiencies and pulse-limited operational bandwidths. With this approach, we characterize the temporal properties of ultrafast laser pulses via nonlinear autocorrelation measurements, leveraging the efficient SHG response of the metasurface to retrieve the pulse duration and temporal profile.
Our results establish 3R-MoS$_2$ as a promising material platform for highly efficient nonlinear nanophotonic applications, and demonstrate the integration of vdW materials in advanced photonic architectures towards compact and scalable devices, including integrated frequency converters, optical signal modulators, and ultrafast pulse measurement technologies. These advances highlight the growing potential of vdW-based nonlinear photonics, paving the way for next-generation optical technologies in information processing, quantum optics, and ultrafast photonics.

\newpage
\noindent
\textbf{Methods}

\noindent
\textbf{3R-MoS$_2$ $\chi^{(2)}$ nonlinear tensor}
The second-harmonic polarization components for 3R-MoS$_2$, belonging to a $3m$ crystal class, generally are given by:
\begin{equation}
\begin{bmatrix}
P_x(2\omega) \\
P_y(2\omega) \\
P_z(2\omega)
\end{bmatrix}
=
2\varepsilon_0
\begin{bmatrix}
0 & 0 & 0 & 0 & \chi_{2} & -\chi_{1} \\
-\chi_{1} & \chi_{1} & 0 & \chi_{2} & 0 & 0 \\
\chi_{3} & \chi_{3} & \chi_{4} & 0 & 0 & 0
\end{bmatrix}
\begin{bmatrix}
E_x(\omega)^2 \\
E_y(\omega)^2 \\
E_z(\omega)^2 \\
2E_y(\omega)E_z(\omega) \\
2E_x(\omega)E_z(\omega) \\
2E_x(\omega)E_y(\omega)
\end{bmatrix}
\end{equation}
\noindent
where we used the following notation: $\chi_{1} \equiv \chi^{(2)}_{yyy} = -\chi^{(2)}_{yxx} = -\chi^{(2)}_{xxy} = -\chi^{(2)}_{xyx}$, $\chi_{2} = \chi^{(2)}_{xzx} = \chi^{(2)}_{yzy} = \chi^{(2)}_{xxz} = \chi^{(2)}_{yyz}$, $\chi_{3} \equiv \chi^{(2)}_{zxx} = \chi^{(2)}_{zyy}$, $\chi_{4} \equiv \chi^{(2)}_{zzz}$. 
For the considered structures these equations can be simplified due to following reasons. When the long side of the metasurface bricks is oriented along $y$ direction the qBIC state is excited with a normally incident $y$-polarized plane wave. Then (i) the dominant component of the total electric field inside the particles is the $y$ component, while two other components are of the order of magnitude less and/or distributed in the lesser volume of the nonlinear material; (ii) $\chi_{2,3,4}$ nonlinear tensor elements are much smaller than $\chi_1$~\cite{zograf2024}; (iii) emissivity of some of the induced nonlinear polarization currents is rather weak due to their specific spatial distribution poorly overlapping with the emitted plane wave. Applying these considerations shows that all terms associated with $\chi_{2,3,4}$ elements result in substantially smaller contributions to the SHG signal compared with the contribution from the $\chi_1$ one ($D_{3h}$ point group), and therefore they can be neglected. Finally, by orienting ZZ and AC crystal axes along \textit{x} and \textit{y} directions, respectively, one obtains Equations (1-2) in the main text. \begin{eqnarray}
P_{ZZ}^{2\omega} &\propto& 2\,E_{AC}\,E_{ZZ}, \\
P_{AC}^{2\omega} &\propto& E_{AC}^2 - E_{ZZ}^2.
\end{eqnarray}
\noindent
The similar consideration applies also in general for arbitrary orientation of the metasurface with respect to the crystal axis, as well as for arbitrary polarization of the incident field.

\noindent
\textbf{Simulations}
All numerical simulations were conducted using the commercial software package COMSOL Multiphysics 6.1. A single unit cell of a periodic structure on an 
SiO$_{2}$ substrate was used as the computational model. To calculate the transmission spectra shown in Figure \ref{fig2}b, the structure was excited normally with a plane wave of appropriate polarization. Excitation and detection ports were positioned at the top and bottom boundaries of the model, respectively. Floquet periodic boundary conditions were applied to the adjacent side faces to account for the periodicity of the structure.
For the SHG calculations, the simulation process was divided into two stages. In the first stage, the structure's response at the fundamental frequency was computed, and the resulting field distribution was obtained. In the second stage, the nonlinear polarization at the second harmonic frequency was introduced, which was derived from the fields obtained in the first stage. The final power, corresponding to the generation spectra displayed in Figure \ref{fig2}e, was determined by integrating the Poynting vector summed over the top and bottom surface of the periodic model.
To investigate the angular dependence of the polarized signal shown in Figure \ref{fig3}e, the polar angle $\phi$ of the incident field was varied according to $E_{AC}=E_{0}\sin(\phi)$ and $E_{ZZ}=E_{0}\cos(\phi)$, where $E_{0}$ is the amplitude of the electric field. Additionally, to model the collection of linearly polarized light, only one component of the nonlinear polarization was considered in the analysis (see also Supplementary Note II).

\noindent
\textbf{Sample fabrication}
Commercial 3R-MoS$_2$ single crystals (HQ Graphene) are exfoliated onto glass substrates. Suitable films are selected using optical microscopy, prioritizing those with large size, uniform thickness, and well-defined crystal axes. The height of the films is measured using a profilometer (Bruker Dektak XT).
The sample is then encapsulated with a flake thickness-dependent SiO$_2$ layer via magnetron sputtering (Angstrom NEXDEP). Next, a PMMA 950k layer is spin-coated and baked on a hotplate at 175 °C for 3 minutes, followed by the application of a discharge layer (Espacer 300Z). The metasurface design is patterned using electron-beam lithography (Raith eLine Plus).
The ebeam resist is developed in an 80\% ethanol solution diluted with water for 20 seconds, and the development process is stopped using iso-propanol. A molecular adhesion layer (MPTMS) is then applied, followed by the deposition of a gold hardmask via electron-beam evaporation. The resist is removed using acetone.
Subsequently, the SiO$_2$ layer is etched using reactive-ion etching with CHF$_3$ chemistry. The gold hardmask is then dissolved using a standard Au etchant (Sigma-Aldrich). Finally, the flake is etched with SF$_6$ chemistry, utilizing the SiO$_2$ layer as a hardmask, which is consumed during the etching process, ultimately yielding the nanostructured 3R-MoS$_2$ flakes.

\noindent
\textbf{Optical spectroscopy}
Optical spectroscopy measurements were performed by mounting the sample in an inverted microscope (Nikon Eclipse). White-light optical transmission spectra of the metasurfaces were acquired using a thermal white-light source (Thorlabs SLS203F). The light was linearly polarized and controlled with a halfwaveplate, and then focused onto the sample through a 10$\times$ objective (NA = 0.25). The transmitted light was collected using a high-NA Nikon Plan Fluor 40$\times$ objective (NA = 0.60) before being analyzed by an AlGaAs infrared spectrometer. For SHG measurements, an optical parametric oscillator (Chameleon Compact OPO-Vis) driven by a Ti:Sapphire laser source (Coherent Chameleon Ultra II) was used, providing a nominal pulse duration of 200 fs. The linearly polarized laser beams are focused onto the sample using the same configuration discussed before. The generated SHG signal was collected through the same high-NA objective. In case of polarization resolved measurements, a combination of a linear polarizer and a waveplate was used to analyze the polarization state of the emitted SHG emission, and the SHG spectrum was recorded using a visible spectrometer.

\bigskip
\noindent
\textbf{Data Availability}

\noindent
The data that support the findings of this study are available at https://doi.org/10.5281/zenodo.XX\\XXXX

\noindent
\textbf{References}
\bibliography{3R}
\bibliographystyle{apsrev4-2}

\bigskip
\noindent
\textbf{Acknowledgements}

\noindent
L.So. thanks Lucas Lafeta for fruitful discussions. This work was funded by the European Union (ERC, METANEXT, 101078018 and EIC, NEHO, 101046329). Views and opinions expressed are however those of the author(s) only and do not necessarily reflect those of the European Union, the European Research Council Executive Agency, or the European Innovation Council and SMEs Executive Agency (EISMEA). Neither the European Union nor the granting authority can be held responsible for them. This project was also funded by the Deutsche Forschungsgemeinschaft (DFG, German Research Foundation) under grant numbers EXC 2089/1–390776260 (Germany’s Excellence Strategy) and TI 1063/1 (Emmy Noether Program), the Bavarian program Solar Energies Go Hybrid (SolTech) and the Center for NanoScience (CeNS). S.A.M. additionally acknowledges the Lee-Lucas Chair in Physics.

\bigskip
\noindent
\textbf{Author Contributions}
L.Se., T.P., and L.So. carried out optical spectroscopy experiments. T.W. fabricated the samples. A.A.S., T.W., and R.S performed the numerical analysis. L.dS.M., A.T., S.A.M, and M.A.P. supervised various aspects of the project. L.So. wrote the manuscript with contributions from all the authors. L.So. and L.dS.M. conceived the idea and oversaw the whole project.

\noindent

\bigskip
\noindent
\textbf{Conflict of interest}

\noindent
The authors declare no competing interests.

\end{document}